\documentclass[a4paper]{amsart}
\usepackage{hyperref} 
\usepackage{fancyhdr}
\usepackage{amsthm}
\usepackage{amsmath}
\usepackage{graphicx}
\usepackage{amssymb}
\makeatletter

\newcommand{\authorfootnotes}{\renewcommand\thefootnote{\@fnsymbol\c@footnote}}%
\makeatother
\begin{document}

\title{A Hierarchical Finite Element Method\\
for Quantum Field Theory}
\maketitle

\begin{center}
 \normalsize
 \authorfootnotes Arnab Kar\footnote{arnabkar@pas.rochester.edu}, Fred Moolekamp, S.~G. Rajeev\footnote{Also at the Department of Mathematics}\par  \bigskip
Department of Physics and Astronomy,\\ 
University of Rochester,\\
Rochester, New York 14627, USA.\\ 
\end{center}
\begin{abstract}
We study a model of scalar quantum field theory in which space-time is a discrete set of points obtained by repeatedly subdividing a triangle into three triangles at the centroid. By integrating out the field variable at the centroid we get a renormalized action on the original triangle. The exact renormalization map between the angles of the triangles is obtained as well. A fixed point of this map happens to be the cotangent formula of Finite Element Method which approximates the Laplacian in two dimensions.
\end{abstract}
\begin{description}
\item[PACS numbers] 02.70.Dh,  41.20.Cv,  11.25.Hf
\end{description}

\section{Introduction}
The most successful regularization method in understanding non-perturbative
Quantum Field Theory (QFT) is the lattice method,\cite{LatticeQFT1,LatticeQFT2} which replaces 
space-time by a periodically arranged finite set of points. Numerical
simulations based on this are becoming increasingly accurate, therefore any
attempt at a mathematical formulation of quantum field theory must
build on this success and aim to improve upon it.

The classical analogue of the problem would be 
the solution of Partial Differential Equations (PDEs). In the early
days a lattice with identically shaped fundamental regions was used
in numerical solutions of PDEs. Later it was realized that using
meshes adapted to the boundary conditions makes more economical use
of computing resources by adding more points where the field varies rapidly
and fewer where it varies slowly. The Finite Element Method\cite{FEM,Sorkin} was developed
in the seventies: it allows fundamental cells to have different shapes
and sizes and use sophisticated interpolation methods to model the
field in the interior of each cell. Some of the mathematical ideas
were anticipated by Whitney\cite{Whitney} in his work in topology
and extended by Patodi\cite{Patodi}.
The Whitney elements have provided a basis for a discrete formulation
of geometry. This Discrete Differential Geometry is useful not only to solve PDEs,
but also to model shapes for use in computer graphics\cite{DDG1,DDG2}.

The analogue in Quantum Field Theory is to replace the periodic lattice
with a mesh that contains different length scales. This approach has been looked at by groups
in the past and met with varying degrees of success. The first 
approach in this direction was by Christ, Friedberg and Lee\cite{RandomLattice} (except
they proposed to average over all locations of lattice points as a
way to restore rotation invariance, which did not turn out to be helpful).
There is also some early work by Bender, Guralnik and Sharp\cite{Benderetal}.
Patodi's FEM to solve the eigenvalue problem for Laplacians was not
noticed by physicists at this time. Since then much of the work on Lattice
Gauge Theories has been computational along with some analytic work\cite{HalvorsenSrensen}.

We propose to adapt existing methods of QFT and develop new Finite Element
Methods to understand the essential problem from Wilson's point of view: how to integrate out
some variables and get an effective theory for the remaining degrees
of freedom (for a recent review, see the volume\cite{Renorm}). The simplest case is the
one dimensional lattice (the set of integers), which has a natural 
subdivision into even and odd numbered elements. By
integrating out the odd sites  and leaving only the even sites we are left with an identical
lattice with a different separation between field points. Unfortunately there is
no simple procedure to extend this into higher dimensions.

A natural idea would be to divide space into triangles (simplices
in higher dimensions) and to fit them together to form larger ones, allowing us to
integrate out the interior vertices and obtain an effective
large scale theory. An advantage of our regularization method is that 
the renormalization map can be calculated exactly.
The transformation between the angles of the 
triangles from subsequent generations is obtained at each stage of the subdivision. 
The first examples\cite{KarRajeevFractals} we
constructed this way ignored the shape (information contained in the
angles) of the triangles. The Finite Element Method 
used by engineers leads to a ``cotangent formula''\cite{Duffin}. It approximates 
the Laplacian in two dimensions on one hand and also happens to 
be a fixed point\cite{Dittrich} of the renormalization dynamics. We determine this dynamics explicitly.

We expect this fixed point to be a continuum limit on a fractal, analogous
to the Bethe lattice for which the renormalization group can be exactly 
calculated. Such QFTs can serve as approximations to theories on Euclidean 
spaces. Or perhaps at short distances, space-time really is not Euclidean.  

If generalized to the Ising model, nonlinear sigma models 
or to four dimensional field theories, we could get interesting examples
of Discrete Conformal Field Theory\cite{DCFT}. In our approach we do not average over triangulations.
Such an average has been proposed as an approach to quantum gravity\cite{cdt} 
and as a way to restore translation invariance\cite{RandomLattice}.

\section{The Cotangent Formula}
In the early days of computational engineering, Duffin\cite{Duffin} derived a formula
for the discrete approximation for the energy of an electrostatic
field on a planar domain. In this Finite Element Method the plane is 
divided into triangles where the field is specified at each vertex and
the energy of the field is the sum of contributions from each triangle.
An approximation for the energy of a triangle is obtained by linear
interpolation of the field to the interior.

Suppose the vertices $x_{0},x_{1},x_{2}$ correspond to
field values $\phi_{0},\phi_{1},\phi_{2}$. Each point $x$ in the
interior of the triangle divides it into three sub-triangles with vertices
$\{x,x_{0},x_{1}\}$, $\{x,x_{1},x_{2}\}$ and $\{x,x_{2},x_{0}\}$
respectively. 

If the ratio of the area of a sub-triangle opposite to $x_{0}$ to the larger triangle is
\[
u_{0}=\frac{\Delta( x,x_1,x_2)}{\Delta (x_0,x_1,x_2)},
\]
then 
\[
x=u_{0}x_{0}+u_{1}x_{1}+u_{2}x_{2},\quad u_{0}+u_{1}+u_{2}=1,\quad u_{0},u_{1},u_{2}>0.
\]

We can use the pair $u_{0},u_{1}$ as co-ordinates instead of the
cartesian components of $x$. The linear interpolation of the field
values to the point $x$ is then
\[
\phi(x)=u_{0}\phi_{0}+u_{1}\phi_{1}+u_{2}\phi_{2}.
\]

The energy of the interpolated field inside a triangle on calculation 
turns out to be 
\[
S=\frac{1}{4}\left[a_{2}\left(\phi_{0}-\phi_{1}\right)^{2}+a_{1}\left(\phi_{2}-\phi_{0}\right)^{2}+a_{0}\left(\phi_{1}-\phi_{2}\right)^{2}\right]
\]
where $a_{0},a_{1},a_{2}$ are the cotangents of the angles at the
vertices.

\begin{figure}
\centerline{\includegraphics[scale=0.5]{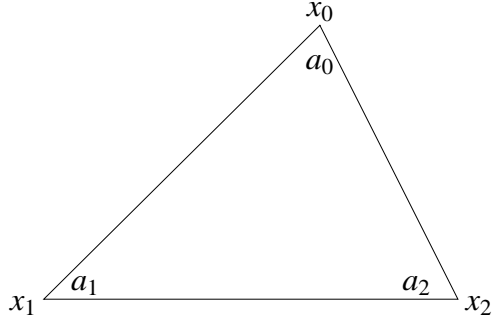}}
\vspace*{8pt}
\caption{A triangle with labelled vertices and cotangents of the angles.\protect\label{fig1}}
\end{figure}
\begin{proof}
Define the vectors along the sides of the triangle (see Fig.~\ref{fig1}),
\[
e_{1}^{\mu}=x_{1}^{\mu}-x_{0}^{\mu},\quad e_{2}^{\mu}=x_{2}^{\mu}-x_{0}^{\mu}.
\]

Using $u^{a}$ for $a=1,2$ as co-ordinates, 
\[
x^{\mu}=u^{a}e_{a}^{\mu}\implies\partial_{a}x^{\mu}=e_{a}^{\mu}
\]

Then the metric tensor of the plane in these co-ordinates has as components
the dot products of the sides:
\[
g_{ab}=e_{a}^{\mu}e_{b}^{\nu}\delta_{\mu\nu},\quad g=\left(\begin{array}{cc}
|e_{1}|^{2} & e_{1}\cdot e_{2}\\
e_{1}\cdot e_{2} & |e_{2}|^{2}
\end{array}\right)
\]

Also, $\sqrt{\det g}=e_{1}\times e_{2}$ is twice the area of the
triangle. The cotangents are 
\[
a_{0}=\frac{e_{1}\cdot e_{2}}{e_{1}\times e_{2}},\quad a_{1}=\frac{(e_{2}-e_{1})\cdot e_{1}}{(e_{2}-e_{1})\times e_{1}},\quad a_{2}=\frac{e_{2}\cdot(e_{2}-e_{1})}{e_{2}\times(e_{2}-e_{1})}
\]

Then, 
\[
a_{0}+a_{1}=\frac{|e_{1}|^{2}}{e_{1}\times e_{2}},\quad a_{0}+a_{2}=\frac{|e_{2}|^{2}}{e_{1}\times e_{2}}
\]
and 
\[
\sqrt{\det g}g^{ab}=\left(\begin{array}{cc}
\frac{|e_{2}|^{2}}{e_{1}\times e_{2}} & -\frac{e_{1}\cdot e_{2}}{e_{1}\times e_{2}}\\
-\frac{e_{1}\cdot e_{2}}{e_{1}\times e_{2}} & \frac{|e_{1}|^{2}}{e_{1}\times e_{2}}
\end{array}\right)=\left(\begin{array}{cc}
a_{0}+a_{2} & -a_{0}\\
-a_{0} & a_{0}+a_{1}
\end{array}\right)
\]

Thus, using $\int d^{2}u=\frac{1}{2}$,
\begin{eqnarray*}
S&=&\frac{1}{2}\int\sqrt{\det g}g^{ab}\partial_{a}\phi\partial_{b}\phi d^{2}\tau\\
&=&\frac{1}{4}\left[(a_{0}+a_{2})(\phi_{1}-\phi_{0})^{2}-2a_{0}(\phi_{1}-\phi_{0})(\phi_{2}-\phi_{0})+(a_{0}+a_{1})(\phi_{2}-\phi_{0})^{2}\right]\\
\end{eqnarray*}

This can be rewritten as
\[
S(\phi_{0},\phi_{1},\phi_{2}|a_{0},a_{1},a_{2})=\frac{1}{4}\left[a_{0}(\phi_{1}-\phi_{2})^{2}+a_{1}\left(\phi_{2}-\phi_{0}\right)^{2}+a_{2}\left(\phi_{0}-\phi_{1}\right)^{2}\right]
\]
as claimed.
\end{proof}

\section{The Geometry of Triangles}
The space $\mathcal{S}$ of similarity classes of triangles (with
marked vertices) is a hyperboloid\cite{BaranyEtAl}. This can be understood in several
ways. A pair of sides of a triangle forms a basis, thus the space
of marked triangles may be identified with $GL(2,\mathbb{R})$: this group
acts transitively and without a fixed point on the space of bases. Quotienting
by rotation, scaling and reflection around a side gives
\[
\mathcal{S}=GL(2,\mathbb{R})/\left(SO(2,\mathbb{R})\times \mathbb{R}^{+}\times \mathbb{Z}_{2}\right)=SL(2,\mathbb{R})/SO(2,\mathbb{R}).
\]
which is a hyperboloid. This argument generalizes to $n$ dimensions:
the similarity classes of marked simplices is $GL(n,\mathbb{R})/\mathbb{R}^{+}\times SO(n,\mathbb{R})\times \mathbb{Z}_{2}=SL(n,\mathbb{R})/SO(n,\mathbb{R})$.

An equivalent point of view is that $\mathcal{S}$ is the space of
symmetric tensors of determinant one: a pair of sides of the triangles
define a symmetric tensor through their inner products. By scaling
we can choose this symmetric tensor to have determinant one. It is
clear that $SL(2,\mathbb{R})$ acts on the space of such tensors transitively,
with $SO(2,\mathbb{R})$ as the isotropy group at one point. Again this generalizes
to $n$ dimensions.

A more explicit point of view will be useful in what follows. A similarity
class of marked triangles is determined by the angles
at the vertices (or, for convenience, the cotangents of the angles). Since the angles $(\theta_0,\theta_1,\theta_2)$ of a triangle add up to $\pi$, the cotangents
satisfy 
\begin{equation}
a_{0}a_{1}+a_{1}a_{2}+a_{2}a_{0}=1,\quad a_i=\cot\theta_i \label{CotangentIdentity}
\end{equation}

This can be written as 
\[
a^{T}\eta a=1,\quad\quad a=\left(\begin{array}{c}
a_{0}\\
a_{1}\\
a_{2}
\end{array}\right),\quad\eta=\frac{1}{2}\left(\begin{matrix}0 & 1 & 1\cr1 & 0 & 1\cr1 & 1 & 0\end{matrix}\right)
\]
Since $\eta$ has signature ($1,-1,-1)$, this is the equation for
a time-like hypersurface in Minkowski space $\mathbb{R}^{1,2}$. Setting 
\[
p_{0}=\frac{a_{1}+a_{2}+a_{0}}{\sqrt{3}},\quad p_{1}=\frac{a_{2}-a_{1}}{2},\quad p_{2}=\frac{2a_{0}-a_{1}-a_{2}}{2\sqrt{3}}
\]
the ``cotangent identity" (\ref{CotangentIdentity}) becomes the equation for a  hyperboloid
\[
p_{0}^{2}-p_{1}^{2}-p_{2}^{2}=1.
\]
The quantity $4\,(a_0+a_1+a_2)$ is the ratio of the sum of squares of the sides to the area of the triangle. It is a minimum for an equilateral triangle and becomes large for a flat triangle (one with small area or large perimeter).

So far we discussed triangles with marked vertices but we should also consider invariant transformations of the vertices. The group $S_{3}$ of permutations of vertices is generated by the
cyclic permutation
\[
\sigma:012\mapsto120
\]
and the interchange of a pair of vertices
\[
\tau:012\mapsto021
\]
\[
S_{3}=\langle\sigma,\tau|\sigma^{3}=1,\tau^{2}=1,\tau\sigma\tau=\sigma^{2}\rangle.
\]

These permutations act on the cotangents through the matrices
\[
\sigma=\left(\begin{array}{ccc}
0 & 1 & 0\\
0 & 0 & 1\\
1 & 0 & 0
\end{array}\right),\quad\tau=\left(\begin{array}{ccc}
1 & 0 & 0\\
0 & 0 & 1\\
0 & 1 & 0
\end{array}\right).
\]

We can also parametrize $\mathcal{S}$ by the complex number 
\[
z=\frac{a_{1}+i}{a_{1}+a_{2}}.
\]

By a translation, we can choose the first vertex $x_{1}=0$ and by a
rotation and scaling we may choose $x_{2}=1$. $z$ is then the co-ordinate
of the remaining vertex.

Then the permutation of the vertices becomes 
\[
\sigma(z)=\frac{1}{1-z},\quad\tau(z)=1-\bar{z}.
\]

By a reflection around the side $12$, we can choose $a_{0}+a_{1}+a_{2}>0$;
equivalently $\mathrm{Im}(z)>0$. Note that $\tau$ is the reflection
around the perpendicular from vertex $0$ to the opposite side $12$  of the triangle.

\subsection{Subdivision of a triangle}
We can subdivide a triangle into three sub-triangles of equal area
by connecting the centroid $x_{3}=\frac{x_{0}+x_{1}+x_{2}}{3}$ to
the vertices $x_{0},x_{1},x_{2}$ by straight lines. (If we subdivide at some other interior point, we get similar results).

\begin{figure}
\centerline{\includegraphics[scale=0.6]{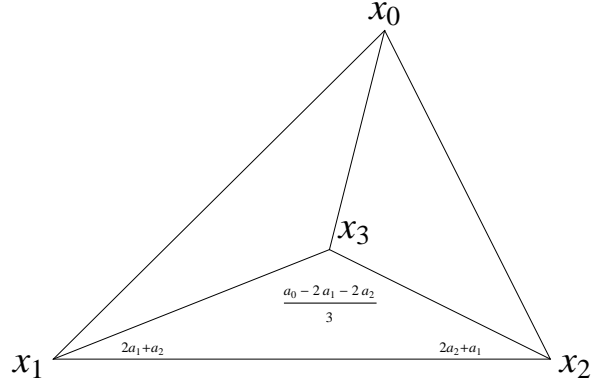}}
\vspace*{8pt}
\caption{A triangle with few cotangents of the angles labelled after subdivision.\protect\label{fig2}}
\end{figure}

The cotangents of the angles of the sub-triangle opposite vertex $0$
are given by 
\[
\cot(x_{2}x_{1}x_{3})=2a_{1}+a_{2},\quad\cot(x_{3}x_{2}x_{1})=2a_{2}+a_{1},\quad\cot(x_{1}x_{3}x_{2})=\frac{a_{0}-2a_{1}-2a_{2}}{3}
\]
as shown in Fig.~\ref{fig2}.

To see this, choose a co-ordinate system with $x_{1}=(0,0),\,x_{2}=(1,0),\,x_{0}=(x,y)$
so that $x_{3}=(\frac{1+x}{3},\frac{y}{3}).$ Then,
\[
a_{1}=\frac{x}{y},\quad a_{2}=\frac{1-x}{y},\quad a_{0}=\frac{1-a_{1}a_{2}}{a_{1}+a_{2}}.
\]

By dropping a perpendicular from $x_{3}$ to the side $x_{1}x_{2}$
we get 
\[
\cot(x_{2}x_{1}x_{3})=\frac{\frac{1+x}{3}}{\frac{y}{3}}=2a_{1}+a_{2},\quad\cot(x_{3}x_{2}x_{1})=\frac{1-\frac{1+x}{3}}{\frac{y}{3}}=2a_{2}+a_{1}.
\]

The remaining angle is given by solving the cotangent formula:
\[
\cot(x_{1}x_{3}x_{2})=\frac{1-(2a_{1}+a_{2})(2a_{2}+a_{1})}{(2a_{1}+a_{2})+(2a_{1}+a_{2})}=\frac{1-a_{1}a_{2}-2(a_{1}+a_{2})^{2}}{3(a_{1}+a_{2})}=\frac{a_{0}-2a_{1}-2a_{2}}{3}.
\]

We can thus express the cotangents of this sub-triangle as $\Lambda a$
where
\[
\Lambda=\left(\begin{array}{ccc}
\frac{1}{3} & -\frac{2}{3} & -\frac{2}{3}\\
0 & 2 & 1\\
0 & 1 & 2
\end{array}\right).
\]

Note that 
\[
\Lambda^{T}\eta\Lambda=\eta
\]
since the cotangent identity is preserved. Thus subdivisions are represented
by Lorentz transformations in $\mathbb{R}^{1,2}$. Note the symmetry under the
interchange of $1$ and $2$:
\[
\Lambda\tau=\tau\Lambda
\]

\begin{figure}
\centerline{\includegraphics[scale=0.5]{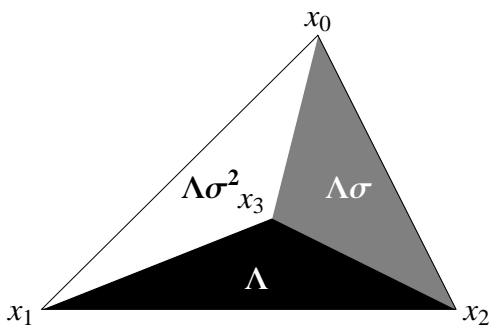}}
\vspace*{8pt}
\caption{Action of $\Lambda$ and $\sigma$ matrices to produce subdivision of a triangle.\protect\label{fig3}}
\end{figure}

The cotangents of the remaining sub-triangles are given by cyclic permutations
$\Lambda\sigma$ and $\Lambda\sigma^{2}$ (see Fig.~\ref{fig3}). In this convention, the
central angle is listed first. 

In the complex parametrization $z=x+iy=\frac{a_{1}+i}{a_{1}+a_{2}}$, the 
subdivision $\Lambda$ corresponds to 
\[
\Lambda(z)=\frac{1+z}{3}
\]
which is the complex co-ordinate of the centroid when $x_{1}=0,\,x_{2}=1,\,x_{0}=z$. Recall
that in this parametrization $\sigma(z)=\frac{1}{1-z}.$ Clearly, both 
$\Lambda$ and $\sigma$ map the upper half plane to itself.

The semi-group generated by $\langle\Lambda,\Lambda\sigma,\Lambda\sigma^{2}\rangle$
describe repeated subdivisions of a triangle. After many iterations, most of the triangles are flat: they have small area and large perimeter\cite{DiaconisMiclo,ButlerGraham}.
The dynamics generated
by this semi-group is the  renormalization
group of real space decimations.

\section{Renormalization Dynamics}
Consider a Gaussian scalar field with values $\phi_{0},\phi_{1},\phi_{2}$
at the vertices of a triangle with cotangents $a_{0},a_{1},a_{2}$.
The most general quadratic form for the discrete approximation to the action will be
\[
S(\phi_{0},\phi_{1},\phi_{2}|a)=P(a)\phi_{0}^{2}+Q(a)\phi_{1}\phi_{2}+P(\sigma a)\phi_{1}^{2}+Q(\sigma a)\phi_{2}\phi_{0}+P(\sigma^{2}a)\phi_{2}^{2}+Q(\sigma^{2}a)\phi_{0}\phi_{1}
\]

The coefficients $P(a),Q(a)$ are functions of the cotangents satisfying
the symmetry
\[
P(a)=P(\tau a),\quad Q(a)=Q(\tau a).
\]

For example, the cotangent formula corresponds to the choice
\[
P(a)=\frac{a_{1}+a_{2}}{4},\quad Q(a)=-\frac{a_{0}}{2}.
\]

If we subdivide the triangle and associate a field $\phi_{3}$ at
the central vertex, the action will be the sum of contributions from
each triangle.
\begin{eqnarray*}
S_{\mathrm{sub}}(\phi_{0},\phi_{1},\phi_{2},\phi_{3}|a)&=&S(\phi_{3},\phi_{1},\phi_{2}|\Lambda a)+S(\phi_{3},\phi_{2},\phi_{0}|\Lambda\sigma a)+S(\phi_{3},\phi_{0},\phi_{1}|\Lambda\sigma^{2}a) \\
&=&A\phi_{3}^{2}+B\phi_{3}+C\\
\end{eqnarray*}
where 
\begin{eqnarray*}
A&=&P(\Lambda a)+P(\Lambda\sigma a)+P(\Lambda\sigma^{2}a)\\
B&=&\phi_{0}\left\{ Q(\sigma\Lambda\sigma a)+Q(\sigma^{2}\Lambda\sigma^{2}a)\right\} +\phi_{1}\left\{ Q(\sigma^{2}\Lambda a)+Q(\sigma\Lambda\sigma^{2}a)\right\} +\phi_{2}\left\{ Q(\sigma\Lambda a)+Q(\sigma^{2}\Lambda\sigma a)\right\}\\
C&=&\phi_{0}^{2}\left[P(\sigma^{2}\Lambda\sigma a)+P(\sigma\Lambda\sigma^{2}a)\right]+\phi_{1}^{2}\left[P(\sigma\Lambda a)+P(\sigma^{2}\Lambda\sigma^{2}a)\right]+\phi_{2}^{2}\left[P(\sigma^{2}\Lambda a)+P(\sigma\Lambda\sigma a)\right]\\
&&+\phi_{1}\phi_{2}Q(\Lambda a)+\phi_{2}\phi_{0}Q(\Lambda\sigma a)+\phi_{0}\phi_{1}Q(\Lambda\sigma^{2}a)\\
\end{eqnarray*}

The effective action after integrating out the central field variable
is given by 
\[
e^{-\tilde{S}(\phi_{0},\phi_{1},\phi_{2}|a)}=Z\int e^{-S_{\mathrm{sub}}(\phi_{0},\phi_{1},\phi_{2},\phi_{3}|a)}d\phi_{3}
\]
where $Z=\sqrt{\frac{3(a_{0}+a_{1}+a_{2})}{2\pi}}$ is a normalization constant.
\[
\tilde{S}(\phi_{0},\phi_{1},\phi_{2}|a)=C-\frac{B^{2}}{4A}
\]
On comparing coefficient of $\phi_0^2$, we get
\[
\tilde{P}(a)=P(\sigma^{2}\Lambda\sigma a)+P(\sigma\Lambda\sigma^{2}a)-\frac{\left[Q(\sigma\Lambda\sigma a)+Q(\sigma^{2}\Lambda\sigma^{2}a)\right]^{2}}{4\left[P(\Lambda a)+P(\Lambda\sigma a)+P(\Lambda\sigma^{2}a)\right]}
\]

On comparing coefficient of $\phi_1\phi_2$, we get
\[
\tilde{Q}(a)=Q(\Lambda a)-\frac{\left[Q(\sigma^{2}\Lambda a)+Q(\sigma\Lambda\sigma^{2}a)\right]\left[Q(\sigma\Lambda a)+Q(\sigma^{2}\Lambda\sigma a)\right]}{2\left[P(\Lambda a)+P(\Lambda\sigma a)+P(\Lambda\sigma^{2}a)\right]}
\]

Using $\Lambda\tau=\tau\Lambda,\tau\sigma\tau=\sigma^{2}$ we can
verify that $\tilde{P}(\tau a)=\tilde{P}(a),\tilde{Q}(\tau a)=\tilde{Q}(a)$
as needed for symmetry. The denominator 
\[
A(a)=P(\Lambda a)+P(\Lambda\sigma a)+P(\Lambda\sigma^{2}a)
\]
is invariant under $\sigma,\tau$ and hence, under all permutations.

The semi-group generated by the map $\mathcal{R}:(P,Q)\mapsto(\tilde{P},\tilde{Q})$
on the space of pairs of functions on the hyperboloid is the renormalization
dynamics (``renormalization group''). This explicit example should help understand such dynamics.
For example, is there is a notion of entropy that increases monotonically?
Its fixed points  correspond to some sort of continuum limit (which
could be fractals\cite{Strichartz,Kigami}).

\section{Fixed Points}
An obvious fixed point consists of constant $P,Q$. This corresponds to the ``Apollonian subdivisions" considered in an earlier paper\cite{KarRajeevFractals}.

We now show that the cotangent formula of the FEM
\[
P(a)=\frac{a_{1}+a_{2}}{4},\quad Q(a)=-\frac{a_{0}}{2}
\]
is also a fixed point\cite{Dittrich} of the above dynamics. It is not hard to verify that 
\begin{eqnarray*}
A=P(\Lambda a)+P(\Lambda\sigma a)+P(\Lambda\sigma^{2}a)&=&\frac{3}{2}(a_{0}+a_{1}+a_{2})\\
Q(\sigma\Lambda\sigma a)+Q(\sigma^{2}\Lambda\sigma^{2}a)&=&-(a_{0}+a_{1}+a_{2})\\
P(\sigma^{2}\Lambda\sigma a)+P(\sigma\Lambda\sigma^{2}a)&=&\frac{1}{12}\left[2a_{0}+5(a_{1}+a_{2})\right]\\
\end{eqnarray*}
so that $\tilde{P}(a)=\frac{a_{1}+a_{2}}{4}$.

Similarly,
\begin{eqnarray*}
Q(\Lambda a)&=&\frac{-a_{0}+2a_{1}+2a_{2}}{6}\\
Q(\sigma^{2}\Lambda a)+Q(\sigma\Lambda\sigma^{2}a)&=&-(a_{0}+a_{1}+a_{2})\\
Q(\sigma\Lambda a)+Q(\sigma^{2}\Lambda\sigma a)&=&-(a_{0}+a_{1}+a_{2})\\
\end{eqnarray*}
from which $\tilde{Q}(a)=-\frac{a_{0}}{2}$ follows.

This fixed point describes some sort of continuum limit of two dimensional scalar field theory. As in the examples of 
Ref.~\cite{KarRajeevFractals} it is likely to be a fractal of dimension less than two; but we have not been able to determine this dimension yet. An extension of this method to higher dimensions and to gauge theories would be interesting. We hope to return to these issues in the future.

\section{Acknowledgement}
We thank  Abdelmalek Abdesselam, Abhishek Agarwal, Alex Iosevich and V.~Parameswaran Nair for discussions related to this work. We also thank Bianca Dittrich for bringing Ref.~\cite{Dittrich} to our attention.

\end{document}